\documentclass{emulateapj}
\usepackage{apjfonts}
\slugcomment{Accepted by \textsc{the Astrophysical Journal Letters}}

\newcommand{\geff}{g_{\mathrm{eff}}}
\newcommand{\sdot}{\dot{\Sigma}}
\newcommand{\sdotcrit}{\dot{\Sigma}_{\mathrm{crit}}}
\newcommand{\mdotcrit}{\dot{M}_{\mathrm{crit}}}
\newcommand{\mdotedd}{\dot{M}_{\mathrm{Edd}}}
\newcommand{\siglayer}{\Sigma_{\mathrm{layer}}}
\newcommand{\sigign}{\Sigma_{\mathrm{ign}}}
\newcommand{\lambdaign}{\lambda_{\mathrm{ign}}}
\newcommand{\tign}{t_{\mathrm{ign}}}

\begin{document}

\title{The Latitude of Type I X-Ray Burst Ignition on Rapidly Rotating 
Neutron Stars}
\shorttitle{Burst Ignition Latitude}
\author{Randall L.\ Cooper and Ramesh Narayan}
\shortauthors{Cooper \& Narayan}
\affil{Harvard-Smithsonian Center for Astrophysics, 60 Garden Street,
Cambridge, MA 02138}

\email{rcooper@cfa.harvard.edu, rnarayan@cfa.harvard.edu}

\begin{abstract}

We investigate the latitude at which type I X-ray bursts are ignited
on rapidly rotating accreting neutron stars.  We find that, for a wide
range of accretion rates $\dot{M}$, ignition occurs preferentially at
the equator, in accord with the work of Spitkovsky et al.  However,
for a range of $\dot{M}$ below the critical $\dot{M}$ above which
bursts cease, ignition occurs preferentially at higher latitudes.  The
range of $\dot{M}$ over which nonequatorial ignition occurs is an
increasing function of the neutron star spin frequency.  These
findings have significant implications for thermonuclear flame
propagation, and they may explain why oscillations during the burst rise
are detected predominantly when the accretion rate is high.  They also
support the suggestion of Bhattacharyya \& Strohmayer that
non-photospheric radius expansion double-peaked bursts and the unusual
harmonic content of oscillations during the rise of some bursts result
from ignition at or near a rotational pole.

\end{abstract} 


\section{Introduction}

Type I X-ray bursts are thermonuclear explosions that occur on the
surfaces of accreting neutron stars in low-mass X-ray binaries (LMXBs)
\citep{Betal75,GH75,Getal76,BCE76,WT76,J77,MC77,LL77,LL78}, and they
are triggered by unstable hydrogen or helium burning \citep[for
reviews, see][]{C04,SB06}.  The physics of type I X-ray bursts is
generally well understood, and detailed time-dependent one-dimensional
models \citep[e.g.,][]{Wetal04,FGWD06} have been rather successful at
reproducing the gross characteristics of bursts, such as their fast
rise times of $\sim 1$ s, durations of $\sim 10$--$100$ s, and
recurrence times of a few hours to days.

Nearly all type I X-ray burst models are one-dimensional and thus
implicitly assume that matter accretes spherically onto a nonrotating
neutron star.  Therefore, ignition occurs simultaneously over the
entire stellar surface.  However, these assumptions are clearly
inapplicable to the vast majority of accreting neutron stars.  The
{\it Rossi X-Ray Timing Explorer} has detected highly sinusoidal
oscillations with frequencies of $45$--$1122$ Hz during bursts in $17$
LMXBs \citep[Strohmayer \& Bildsten 2006 and references
therein;][]{BSMS06,B06,Ketal07}.  It is thought that the burst
oscillation frequency corresponds to the neutron star spin frequency
\citep[e.g.][]{SM99}, which implies that many neutron stars that
exhibit bursts are in fact rapidly rotating.  Indeed,
\citet{MCGS01,MGC04} have shown that the properties of bursts depend
sensitively on the neutron star rotation rate, and so rotation must be
considered in theoretical burst models.  Furthermore, since the time
required to accumulate a sufficient amount of fuel to trigger a burst
is much greater than the duration of the resulting burst, it is highly
unlikely that all of the accreted fuel over the entire neutron star
surface will ignite simultaneously \citep{J78,S82}.  Thus, contrary to
the assumptions of most theoretical models, ignition almost surely
occurs at a point, and the resulting thermonuclear flame subsequently
engulfs the whole stellar surface in $\sim 1$ s, the burst rise time
\citep[][hereafter SLU02]{FW82,B95,SLU02}.  Time-resolved spectroscopy
and observations of large amplitude oscillations during the rise of
some type I X-ray bursts from rapidly rotating neutron stars support
this conclusion \citep{SZS97,SZSWL98}.

SLU02 investigated the dynamics of localized ignition on a rapidly
rotating neutron star.  They found that the latitude at which
localized ignition occurs is of great importance in determining the
burning front propagation speed.  It can also affect the stability and
nature of zonal flows during the decay phase of bursts \citep{C05b},
the harmonic content of burst rise oscillations \citep{BS05}, and the
light curves of non-photospheric radius expansion (PRE) bursts
\citep{BS06a,BS06b}.  SLU02 assert that, due to the reduction of the
effective gravitational acceleration, ignition is likely to occur at
the equator of a rapidly rotating neutron star.  However, recent
observations suggest otherwise.  \citet{BS05,BS06a,BS06b} argue that
the harmonic content of burst rise oscillations and non-PRE
double-peaked bursts from the LMXB 4U 1636--536, which harbors a
neutron star with spin frequency $\nu = 581$ Hz, require ignition to
oftentimes occur at or near the rotational pole, at least in this
source.

In this Letter, we evaluate theoretically the latitude at which type I
X-ray bursts on rapidly rotating neutron stars are most
probably ignited.  In \S \ref{iglat} we outline the physics that
governs the latitude of type I X-ray burst ignition and determine the
ignition latitude as a function of the global accretion rate.  We
discuss our results and conclude in \S \ref{discussion}.

\section{Ignition Latitude}\label{iglat}

In this section, we first outline the basic physics that determines
the latitude of type I X-ray burst ignition on a rapidly rotating
neutron star using a simple model.  We then use the global linear
stability analysis of \citet{CN05} to carry out more detailed
calculations.

\subsection{Basic Physics}\label{basicphysics}

Using the one-zone burst ignition model of \citet{B98}, we expand upon
the work of SLU02 to determine the most probable latitude of burst
ignition on the surface of a rapidly rotating neutron star.  The
reader is encouraged to refer to these works for further details.  We
neglect general relativistic corrections throughout for clarity.
Following SLU02, we assume that at all times prior to ignition the
accreted plasma is in hydrostatic equilibrium and is at rest in the
corotating frame \citep[e.g.][]{IS99}.  We presume that the accreted
matter spreads in such a way as to minimize the gravitational
potential energy of the accreted layer.  The base of the accreted
layer is therefore an equipotential surface, and so the pressure at
the base of the accreted layer is the same at every latitude
\citep[e.g.,][]{C83}.  Hydrostatic equilibrium thus implies that $P =
\siglayer(\lambda) \geff(\lambda)$ is independent of latitude, where
$P$ is the pressure at the base of the accreted layer, $\siglayer$ is
the column depth, $\geff$ is the effective gravitational acceleration,
and $\lambda \equiv \pi/2-\theta$ is the latitude.  Thus, $\siglayer
\propto \geff^{-1}$, and so
\begin{equation}\label{sdotpropeqn}
\sdot \propto \geff^{-1},
\end{equation}
where $\sdot = \sdot (\lambda)$ is the local mass accretion rate per
unit area.  According to equation (20) of \citet{B98}, the column
depth at which helium ignites $\sigign$ varies as
\begin{equation}\label{sigigneqn}
\sigign \propto \sdot^{-1/5}\geff^{-2/5}.
\end{equation}
Combining equations (\ref{sdotpropeqn}) and (\ref{sigigneqn}) gives
the ignition time as a function of latitude
\begin{equation}\label{tigneqn}
\tign(\lambda) \equiv \sigign/\sdot \propto \geff^{4/5}.
\end{equation}
Thus $\tign$, the time it takes to accrete a critical amount of fuel
to trigger a type I X-ray burst, is an increasing function of $\geff$.
Due to centrifugal acceleration, $\geff$ is lowest at the equator of a
rapidly rotating neutron star, which means that $\tign(\lambda)$ is a
minimum at the equator.  Therefore, ignition will occur preferentially
at the equator (SLU02).

\begin{figure}
\epsscale{1.2}
\plotone{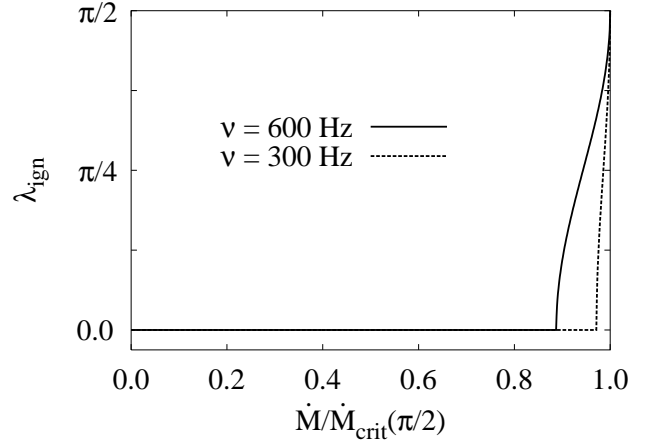}
\caption{Plot shows $\lambdaign$, the latitude at which type I X-ray
bursts ignite on a rapidly rotating neutron star, derived from a
one-zone burst model for two stellar rotation frequencies $\nu$.  $M =
1.4 M_{\odot}$ and $R = 10$ km.  For a wide range of global accretion
rates $\dot{M}$, ignition occurs at the equator ($\lambda = 0$).
However, for a range of $\dot{M}$ near the critical rate above which
bursts cease, ignition occurs at higher latitudes.  The range of
$\dot{M}$ over which nonequatorial ignition occurs is $\propto
\nu^{2}$.
}
\label{onezonefig}
\end{figure}

However, nuclear burning on the surface of an accreting neutron star
does not always trigger a burst.  Both theoretical models
\citep[e.g.,][]{FHM81,P83,B98} and observations
\citep[e.g.][]{Cetal03,Getal06} suggest that there is a local critical
accretion rate $\sdotcrit$ above which nuclear burning is stable, and
thus bursts do not occur.  By equation (\ref{sdotpropeqn}),
$\sdotcrit$ is related to the global critical accretion rate $\dot{M}$
such that
\begin{equation}
\sdotcrit \propto \geff^{-1}\mdotcrit,
\end{equation}
where $\mdotcrit = \mdotcrit(\lambda)$ is defined to be the critical
global accretion rate above which nuclear burning is stable at
latitude $\lambda$.  According to equation (24) of \citet{B98},
\begin{equation}
\sdotcrit \propto \geff^{1/2},
\end{equation}
and so
\begin{equation}\label{mdotgeffeqn}
\mdotcrit \propto \geff^{3/2}.
\end{equation}
This implies that, since $\geff$ is lowest at the equator and
increases towards the rotational pole, there exists a range of
accretion rates for which nuclear burning is stable near the equator
and unstable near the pole.  Consequently, for a given $\lambda_{0}$
and corresponding critical accretion rate $\mdotcrit(\lambda_{0})$,
nuclear burning is stable for all $\lambda < \lambda_{0}$ and unstable
for all $\lambda > \lambda_{0}$.  Clearly, at these accretion rates,
bursts will be triggered off the equator.  For $\dot{M} >
\mdotcrit(\pi/2)$, no bursts are triggered at any latitude.  Taking
the functional derivative of equation (\ref{mdotgeffeqn}) gives
\begin{equation}\label{deltamdot}
\frac{\delta \mdotcrit}{\mdotcrit(\pi/2)} \approx 0.11
\left(\frac{\nu}{600\, \mathrm{Hz}}\right)^{2} \left(\frac{1.4\,
M_{\odot}}{M}\right) \left(\frac{R}{10\, \mathrm{km}}\right)^{3},
\end{equation}
where $\nu$, $M$, and $R$ are the spin frequency, mass, and radius of
the neutron star, respectively.  Thus, for a neutron star with spin
frequency $\nu = 600\,\mathrm{Hz}$, type I X-ray bursts will ignite at
the equator for $\dot{M}/\mdotcrit(\pi/2) \lesssim 0.89$ and off of
the equator for $0.89 \lesssim \dot{M}/\mdotcrit(\pi/2) < 1$.  Within
the latter range of accretion rates, the latitude at which ignition
occurs is given by
\begin{equation}
 \lambdaign = \left \{
\begin{array}{lll}
0, & \dot{M} < \mdotcrit(0),\\ \arccos \left
[\frac{\nu_{\mathrm{K}}}{\nu}
\sqrt{1-\left(\frac{\mdotcrit(\lambda)}{\mdotcrit(\pi/2)}\right)^{2/3}}\right],
& \mdotcrit(0) < \dot{M} < \mdotcrit(\pi/2),\\
\end{array} \right.
\end{equation}
where $\nu_{\mathrm{K}}$ is the Keplerian frequency.  Figure
\ref{onezonefig} shows a plot of $\lambdaign$ as a function of
$\dot{M}$.  We see that type I X-ray burst ignition on a rapidly
rotating neutron star occurs preferentially near the equator for a
wide range of $\dot{M}$.  However, for a small range of $\dot{M}$ near
$\mdotcrit$, ignition occurs at higher latitudes because nuclear
burning becomes stable near the equator.  As equation
(\ref{deltamdot}) illustrates, the range of accretion rates over which
nonequatorial ignition occurs increases dramatically with $\nu$.

\subsection{Global Linear Stability Analysis}

We now use the global linear stability analysis of \citet{CN05}, which
is an expanded and improved version of the model of \citet{NH03}, to
determine the latitude of type I X-ray burst ignition on rapidly
rotating neutron stars.  We assume that matter accretes at a global
rate $\dot{M}$ onto a rapidly rotating neutron star of mass $M =1.4
M_{\odot}$, radius $R = 10$ km, and spin frequency $\nu = 650$ Hz, and
it spreads over the stellar surface in the same manner as described in
\S \ref{basicphysics}.  We set the composition of the accreted matter
to be that of the Sun, such that at the neutron star surface the
hydrogen mass fraction $X = 0.7$, helium mass fraction $Y= 0.28$, CNO
mass fraction $Z_{\mathrm{CNO}} = 0.016$, and heavy element fraction
$Z = 0.004$, where $Z$ refers to all metals other than CNO.

\begin{figure}
\epsscale{1.2}
\plotone{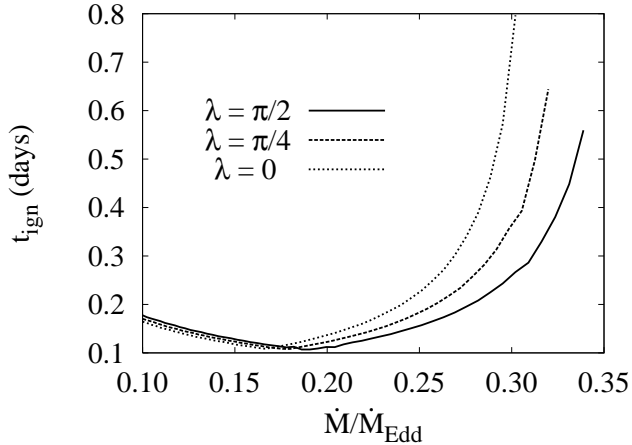}
\caption{Plot shows $\tign(\lambda)$, the type I X-ray burst ignition
timescale, at three different latitudes $\lambda$ for a neutron star
with spin frequency $\nu = 650$ Hz, $M = 1.4 M_{\odot}$, and $R = 10$
km.  For accretion rates $\dot{M} \lesssim 0.16 \mdotedd$, $\tign$ is
lowest at the equator, which means that bursts ignite at the
equator.  For $0.16 \lesssim \dot{M}/\mdotedd \lesssim 0.19$, the
burst ignition latitude increases with $\dot{M}$.  For $0.19 \lesssim
\dot{M}/\mdotedd \lesssim 0.3$, although nuclear burning is unstable
at all $\lambda$, ignition occurs preferentially near the pole where
$\tign$ is lowest.  For $0.3 \lesssim \dot{M}/\mdotedd
\lesssim 0.34$, nuclear burning becomes stable at low latitudes, and
ignition still occurs preferentially near the pole.
}
\label{stabanalysisfig}
\end{figure}

We plot the ignition time $\tign$ as a function of $\dot{M}$ for
$\lambda = 0$, $\pi/4$, and $\pi/2$ in Figure \ref{stabanalysisfig}.
For $\dot{M} \lesssim 0.16 \mdotedd$, $\tign(\lambda)$ is a minimum at
the equator, and so bursts ignite preferentially near the equator, in
agreement with both SLU02 and equation (\ref{tigneqn}).  Here,
$\mdotedd = 8.3 \times 10^{17}$ $\mathrm{g\,s}^{-1}$ denotes the mass
accretion rate at which the accretion luminosity is equal to the
Eddington limit.  For $0.16 \lesssim \dot{M}/\mdotedd \lesssim 0.19$,
$\lambdaign$ gradually increases with increasing $\dot{M}$, again in
accord with the results of \S \ref{basicphysics}.  However, Figure
\ref{stabanalysisfig} shows a new regime of bursts for $\dot{M}
\gtrsim 0.19 \mdotedd$, the regime of ``delayed mixed bursts,'' which
was first identified by \citet{NH03}.  Delayed mixed bursts are mixed 
hydrogen and helium bursts triggered by unstable helium burning, and 
they are preceded by a long period of stable nuclear burning.  
Using their global linear
stability analysis, Narayan \& Heyl found that, within this range of
accretion rates, (1) the ignition timescale is an increasing function
of $\dot{M}$ \citep[which is consistent with observations,
e.g.,][]{vPCLJ79,vPPL88,Cetal03,RLCN06,Getal06}, and (2) the critical
global accretion rate above which delayed mixed bursts occur is an
increasing function of the gravitational acceleration $\geff$.  Thus,
for the range $0.19 \lesssim \dot{M}/\mdotedd \lesssim 0.3$, although
nuclear burning is unstable at all latitudes, ignition occurs
preferentially near the pole where $\tign$ is lowest.  For $0.3
\lesssim \dot{M}/\mdotedd \lesssim 0.34$, nuclear burning becomes
completely stable at low latitudes, and ignition again occurs
preferentially near the pole.  For $\dot{M}/\mdotedd \gtrsim 0.34$,
nuclear burning is stable at all latitudes, and no bursts occur.

\section{Discussion and Conclusions}\label{discussion}

The results of both the simple one-zone model of \citet{B98} and the
more detailed global linear stability analysis of \citet{CN05} suggest
that bursts ignite on the equator at low accretion rates and off of
the equator at higher accretion rates.  The global linear stability
analysis predicts that nonequatorial ignition occurs over a much
larger range of $\dot{M}$ -- nearly $50 \%$ for $\nu = 650$ Hz versus
only about $10 \%$ in the one-zone model -- and that ignition should
occur near the pole for most of this range.
%whereas the one-zone model predicts that $\lambdaign$ is a gradually
%increasing function of $\dot{M}$.  
The differences in the results of the two models are due to the
delayed mixed burst regime, which is predicted in the global linear
stability analysis but not in the one-zone model \citep[for an
explanation of delayed mixed bursts, see][]{CN06}.  In effect, the
delayed mixed burst regime significantly extends the range of
$\dot{M}$ over which nonequatorial ignition occurs.  The global linear
stability analysis and the two-zone model of \citet{CN06} agree better
with observations than all other current burst models, but this
agreement holds only in a time-averaged sense \citep[e.g.][]{Getal06}.
By this we mean that, while observations imply that the mean ignition
time $\langle \tign \rangle$ is an increasing function of $\dot{M}$
for $\dot{M} \gtrsim 0.15 \mdotedd$, which is in accord with the
results shown in Figure \ref{stabanalysisfig}, the ignition time
$\tign$ measured between pairs of bursts observed in nature exhibits
significant deviations relative to the mean.  Our models cannot
account for this behavior.  Since the predicted burst ignition
latitude is model-dependent, and no current model can successfully
reproduce the chaotic behavior of bursts, we have little confidence in
the accuracy of our predicted ignition latitudes $\lambdaign$.
However, all theoretical models would predict that bursts ignite
preferentially near the equator at low values of $\dot{M}$ and off of
the equator at higher $\dot{M}$.  This result is very robust.

Recent observations have generated a renewed interest in the latitude
of type I X-ray burst ignition.  First, non-PRE double peaked bursts
have been observed in a few LMXBs
\citep[e.g.][]{Setal85,Petal89,KHvdKLM02}.  \citet{BS06a,BS06b} argue
that ignition near a rotational pole can explain the non-PRE double
peaked bursts observed in 4U 1636--536, which contains a neutron star
with $\nu = 581$ Hz.  Their simple model of nonequatorial ignition and
subsequent thermonuclear flame propagation and temporary stalling near
the equator qualitatively reproduces both the light curves and spectral
profiles of such bursts.  Second, although burst oscillations detected
in both the rise and decay are usually quite sinusoidal, \citet{BS05}
report evidence for substantial harmonic content in the oscillations
during the rise of a burst from 4U 1636--536.  They again suggest that
nonequatorial ignition and subsequent flame propagation can explain
the observations.  These authors acknowledged that nonequatorial
ignition is at odds with the work of SLU02, who argued for equatorial
ignition at all $\dot{M}$.  Our result that nonequatorial ignition is
more likely when the accretion rate is high naturally reconciles this
discrepancy.

That nonequatorial ignition is preferred at higher $\dot{M}$ may
explain why oscillations in the rising phase of some bursts on rapidly
rotating neutron stars occur predominantly on the banana branch,
i.e. when the inferred accretion rate is high \citep{SB06}.  This can
be understood as follows.  SLU02 find that burning fronts propagate
much faster near the equator than near the poles.  If $\dot{M}$ is low
and ignition occurs near the equator, the flame will spread in
longitude on a timescale that is much shorter than the burst rise
timescale and quickly produce an axisymmetric belt.  An observer would
detect no oscillations during the rising phase because there is no
longitudinal asymmetry.  On the other hand, if $\dot{M}$ is high and
ignition occurs close to, but not directly at, a rotational pole, the
flame will spread in the longitudinal direction on a timescale that is
comparable to the burst rise timescale.  This slow longitudinal spread
would create a non-axisymmetric hot spot and hence oscillations.  We
note that oscillations in the decay phase of some bursts are again
observed only when a system is on the banana branch.  However, burst
oscillations in the decay phase are not well understood theoretically,
and we are hesitant to suggest a connection between these oscillations
and nonequatorial ignition.

We stress that our results apply only to rapidly rotating,
weakly-magnetic neutron stars.  Rapid rotation induces a strong
latitudinal dependence on the effective gravitational acceleration
which restricts burst ignition to a preferred latitude.  For a
nonrotating (or slowly rotating) neutron star, all physical quantities
are independent of $\lambda$ (or nearly so), and ignition could occur
at any latitude.  Thus, oscillations during the burst rise could occur
at any $\dot{M}$ for slowly rotating neutron stars.

Strong magnetic fields may channel and confine accreted matter to some
region of the neutron star surface.  In this case, ignition is more
likely to occur within this confinement region than at a specific
latitude.  Although some bursts observed from the accreting
millisecond pulsar SAX J1808.4--3658 exhibit timing features that
\citet{BS06c} suggest originate from midlatitudinal ignition, we
speculate that it is due to magnetic confinement.

\acknowledgments

We thank Josh Grindlay for discussions that motivated this
investigation and the referee for useful comments.  This work was supported by NASA grant NNG04GL38G.

\bibliographystyle{apj}
%\bibliography{../sbref}

\end{document}